\documentclass[showpacs,amsmath,amssymb,
twocolumn,
aps,prb]{revtex4}
\usepackage{graphicx}

\begin{document}

\title{Electron spin interferometry using a semiconductor ring structure}
\author{Y. K. Kato}
\author{R. C. Myers}
\author{A. C. Gossard}
\author{D. D. Awschalom}
\affiliation{Center for Spintronics and Quantum Computation,University of California, Santa Barbara, CA 93106}
\date{\today}
\begin{abstract}
A ring structure fabricated from GaAs is used to achieve interference of the 
net spin polarization of conduction band electrons. Optically polarized 
spins are split into two packets by passing through two arms of the ring in 
the diffusive transport regime. Optical pumping with circularly polarized 
light on one arm establishes dynamic nuclear polarization which acts as a 
local effective magnetic field on electron spins due to the hyperfine 
interaction. This local field causes one spin packet to precess faster than 
the other, thereby controlling the spin interference when the two packets 
are combined. 
\end{abstract}
\maketitle

Recent progress in electron spin manipulation using non-magnetic 
semiconductors includes the ultrafast all-optical scheme 
\cite{Gupta:2001}, electrical control using g-factor engineering in 
parabolic \cite{Salis:2001,Kato:2003} and coupled quantum wells 
\cite{Poggio:2004}, the strain induced spin-orbit interaction 
\cite{Kato:2004,Kato:2005}, and the spin Hall effect 
\cite{Kato:2006}, demonstrating the broad scope of techniques that can be 
achieved using state-of-the-art semiconductor engineering. The flexibility 
offered by semiconductor spintronics 
\cite{Wolf:2001,Semiconductor:2002,Zutic:2004} is 
anticipated to lead to novel devices and may eventually become useful for 
quantum information processing. Another advantage offered by spin systems in 
semiconductors is their long coherence times. For example, conduction 
electron spins in $n$-type GaAs can have a coherence time exceeding 100~ns and 
can be transported over distances exceeding 100~$\mu $m. 
\cite{Kikkawa:1998,Kikkawa:1999} In contrast, the coherence time 
of the orbital part of the electron wave function is at most a few 
picoseconds even in high-mobility two dimensional systems. Here, we 
demonstrate a device which takes advantage of the long coherence time of the 
carrier spin system. A ring structure is fabricated from $n$-GaAs in which 
electron spins are optically initialized, split into two different paths, 
and recombined on the opposite side. Local nuclear polarization gives rise 
to an additional spin precession phase in one path, causing constructive and 
destructive interference between the two spin packets.

On a semi-insulating (001) GaAs substrate, 2~$\mu $m of undoped 
Al$_{0.4}$Ga$_{0.6}$As and 2~$\mu $m of $n$-GaAs (Si-doped for $n = 3\times 
10^{16}$~cm$^{ - 3})$ are grown by molecular beam epitaxy. The 
interferometer is fabricated from the $n$-GaAs film, while the AlGaAs film 
underneath acts as an etch stop layer. The substrate is polished to $\sim 
$200~$\mu $m prior to processing, and the ring structure [Fig. 1(a)] is 
defined by standard photolithography techniques. The mesa is formed by 
selective spray etching \cite{Tanobe:1992} with a mixture of one part 
NH$_{4}$OH (30{\%}) to 30 parts H$_{2}$O$_{2}$ (35{\%}), and a second 
photolithography step is performed to define the contact areas. The metal 
layers for the contacts are deposited by electron beam evaporation in the 
following order: Ni (5~nm) / Ge (25~nm) / Au (65~nm) / Ni (20~nm) / Au 
(200~nm). The sample is annealed at 420\r{ }C for one minute to form ohmic 
contacts, and a third photolithography step defines a square window on the 
back side of the substrate. Selective spray etching is used again to etch 
the substrate from the back, forming a membrane to allow optical 
transmission experiments. The device has a two-contact resistance of 
6.2~k$\Omega $ at a temperature $T = 5$~K.
\begin{figure}[b]\includegraphics{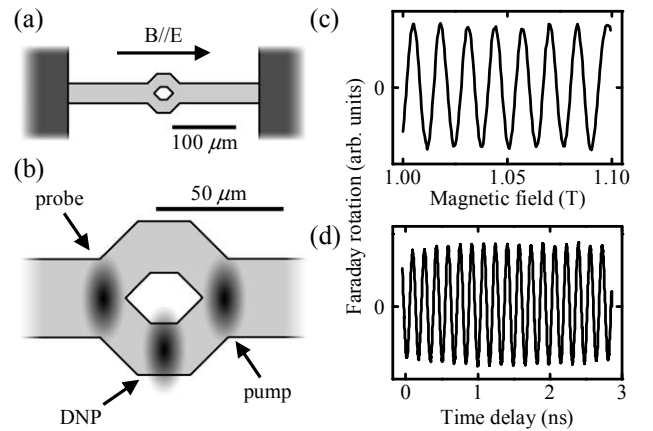}\caption{\label{fig1}
(a) Device schematic. Dark areas are contacts and light gray area is the 
GaAs spin interferometer. (b) Schematic of experimental geometry. (c) 
Faraday rotation as a function of $B$ at $\Delta t = - 10$~ps and an applied 
voltage of 3.5~V. (d) Faraday rotation as a function of $\Delta $t at $B = 
1$~T and an applied voltage of 3.5~V.
}\end{figure}

To monitor the electron spin dynamics in the device, we employ time-resolved 
Faraday rotation \cite{Crooker:1995,Crooker:1997} in the Voigt 
geometry with the sample growth axis parallel to the optical axis. A 
mode-locked titanium sapphire laser produces $\sim $150~fs pulses at a 
repetition frequency of 76~MHz and its wavelength is tuned to 818~nm to 
address the band gap of GaAs. A circularly polarized pump pulse injects spin 
polarized electrons, and the Faraday rotation of a linearly polarized probe 
pulse measures the electron spin component along the laser propagation 
direction at a time delay $\Delta $t. The laser beams are focused to a spot 
size of $\sim $30 $\mu $m, and the average laser powers are 500~$\mu $W and 
60~$\mu $W for the pump beam and the probe beam, respectively. The circular 
polarization of the pump beam is modulated with a photoelastic modulator at 
50~kHz for lock-in detection and $\Delta $t is controlled with a mechanical 
delay line. Measurements are conducted at $T = 5$~K where the longest 
electron spin lifetimes have been observed. \cite{Kikkawa:1998}

We focus the pump beam to the right side of the ring to generate electron 
spin polarization, while the probe beam detects the spins at the left side 
of the ring [Fig. 1(b)]. A positive voltage is applied to the contact on the 
left while the right contact is grounded, establishing an electric field 
which causes the spins to drift across the structure and into the probe 
spot. In general, the Faraday rotation signal can be described as
\begin{equation}
\sum\nolimits_{n = 0}^\infty {S_0 e^{ - \left( {\Delta t + nt_{rep} } 
\right) / \tau }\cos \left[ {\omega _L \left( {\Delta t + nt_{\text{rep}} } 
\right)} \right]} ,
\end{equation}
where $S_{0}$ is the amplitude, $n$ is an integer specifying successive pulses, 
$t_{\text{rep}} = 13.15$~ns is the repetition time of the laser, $\tau $ is 
the spin lifetime, and $\omega _L = g\mu _B B / \hbar $ is the electron 
Larmor frequency, $g$ is the effective electron g factor, $\mu _{B}$ is the 
Bohr magneton, $B$ is the applied magnetic field, and $\hbar $ is the Planck 
constant. \cite{Kikkawa:1998} In order to simplify the data analysis, 
the voltage is chosen such that only the $n = 1$ pulse contributes to the 
signal. This is possible as the spin packets from successive pulses are 
spatially well separated after $t_{rep}$ if a large enough electric field is 
applied. \cite{Kikkawa:1999} In Fig. 1 (c), the magnetic field 
dependence of the Faraday rotation around $B = 1$~T with a bias voltage of 
3.5~V is shown. The absence of other harmonic components indicates that 
pulses with $n > 1$ are not contributing to the signal since the spins 
generated from earlier pulses have drifted past the probe spot. In Fig. 1 
(d), Faraday rotation is plotted as a function of $\Delta t$. We see no 
abrupt jump at $\Delta t = 0$~ns, showing that there is no contribution from 
the $n = 0$ pulse. This is expected as the pump and the probe spots are 
spatially separated. The frequency of the spin precession signal is used to 
extract the electron g-factor, and gives $\left| g \right| = 0.42$. We also 
note that the voltage has been tuned such that the spin precession signal 
has uniform amplitude throughout the available range of $\Delta $t, meaning 
that the center of the spin packet goes through the center of the probe spot 
at around $\Delta t = 1.5$~ns. The strain-induced effective magnetic field 
\cite{Kato:2004} plays a negligible role at these large applied 
magnetic fields.

In order to establish a phase difference between the two paths, optically 
pumped dynamic nuclear polarization (DNP) is utilized. A circularly 
polarized third beam with an average power of 5 mW is derived from the same 
laser and is focused on the lower arm [Fig. 1 (b)]. A slight tilt of the 
sample causes diffraction of the DNP beam and results in some electron spin 
polarization along $B$. A part of the electron spin angular momentum is 
transferred to nuclear spins, establishing DNP along the applied magnetic 
field which acts as an additional effective magnetic field for the electron 
spins through the contact hyperfine interaction. 
\cite{Optical:1984,Salis:2002} In this manner, the electrons 
traveling through the lower arm gain an additional phase to their spin 
precession. After the two packets have recombined, the expected 
time-resolved Faraday rotation is
\begin{equation}
A_u \cos \left( {\omega _L \Delta t} \right) + A_l \cos \left( {\omega _L 
\Delta t + \Phi } \right) = A\cos \left( {\omega _L \Delta t + \phi } 
\right),
\end{equation}
where
\begin{equation}
A = \left( {A_u ^2 + 2A_u A_l \cos \Phi + A_l ^2} \right)^{1 / 2}
\end{equation}
and
\begin{equation}
\phi = \tan ^{ - 1}\left[ {\frac{A_l \sin \Phi }{A_u + A_l \cos \Phi }} 
\right].
\end{equation}
Here, $A_{u}$ and $A_{l}$ are the amplitudes of the spin packets in the upper 
and the lower arms, respectively, $\Phi $ is the phase difference between 
the two packets, and lastly, $A$ and $\phi $ are the amplitude and the phase, 
respectively, of the combined packet. As $\Phi $ is varied, a change in the 
amplitude of the spin precession signal should occur as a result of the 
interference term $A_u A_l \cos \Phi $.

We initialize the device by waiting for 30 minutes at $B = 1$~T with the DNP 
beam on the sample and the bias voltage set to 0~V. The voltage is turned 
off in an effort to localize the nuclear polarization, while the magnetic 
field is applied to increase the nuclear polarization. 
\cite{Kikkawa:2000} After the nuclear polarization has built up, the DNP 
beam is blocked, the voltage is set to 3.5~V, and measurements of Faraday 
rotation as a function of $\Delta $t are made repeatedly [Fig. 2 (a)]. Since 
the DNP beam is turned off, the nuclear spins begin to relax over a 
timescale of $\sim $30 minutes and reduce the phase difference between the 
two paths. As a consequence, the recombined electron spin polarization at 
the left side of the ring cycles through constructive and destructive 
interferences, which manifests as an oscillation in the amplitude of the 
spin precession signal. Two and a half oscillations are observed in 1500~s, 
corresponding to a difference in an average effective magnetic field of 
28~mT between the two arms.
\begin{figure}\includegraphics{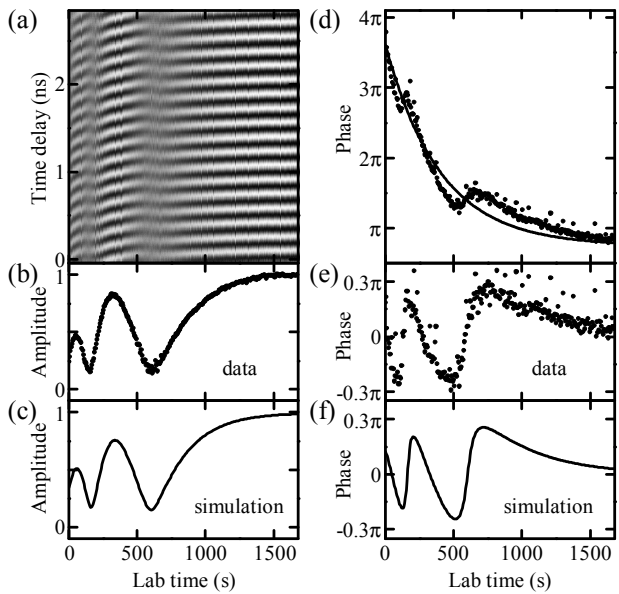}\caption{\label{fig2}
(a) A series of time-resolved Faraday rotation data as a function of lab 
time with an applied voltage of 3.5~V at $T = 5$~K and $B = 1$~T. (b) 
Amplitude of the spin precession signal obtained from fits to data in (a), 
normalized to the value at $t = 1500$ s. (c) Amplitude of the expected spin 
precession signal from the simulation. (d) Phase of the spin precession 
signal from fits to data (filled circles) and exponentially decaying 
background (line). (e) The data in (d) with the background subtracted. (f) 
Phase of the expected spin precession signal from the simulation.
}\end{figure}

In order to quantitatively characterize the interference, we fit individual 
time-resolved Faraday rotation data with $A\cos \left( {\omega _L \Delta t + 
\phi } \right)$. The parameters $A$ and $\phi $ obtained from the fits are 
plotted as a function of lab time in Fig. 2(b) and (d), respectively. The 
amplitude does not dip down to zero, which is expected if the spin packets 
from the two arms do not have equal amplitudes. Additionally, the 
oscillation amplitude increases with lab time, which we attribute to the 
decrease in signal at early times due to inhomogeneous nuclear polarization 
that extends to both arms. Diffusion of spin polarized electrons generated 
by the DNP beam can result in such nuclear polarization, which in turn 
causes the electron spins used for the interference to dephase. The 
dephasing will diminish as the nuclear spins depolarize, and this will 
increase the amplitude of the spin precession signal. This is consistent 
with the behavior of the phase, which shows an exponential decay with small 
oscillations superimposed. This decay can be explained by changes in $\omega 
_{L}$ as a result of nuclear polarization in both arms. 

Assuming that both the amplitude and the phase difference recover with the 
nuclear spin relaxation time $\tau _{n}$, the amplitude of each packet and 
the phase difference are modeled as 
\begin{equation}
A_u = P_u \left( {1 - p_u e^{ - t / \tau _n }} \right),
\end{equation}
\begin{equation}
A_l = P_l \left( {1 - p_l e^{ - t / \tau _n }} \right),
\end{equation}
and
\begin{equation}
\Phi = \Phi _0 e^{ - t / \tau _n }.
\end{equation}
Here, $P_u $ and $P_l = 1 - P_u $ are the fractions of upper and lower arms 
contributing to signal, respectively, $p_u $ and $p_l $ represent the 
portions of the signal which have been reduced due to inhomogeneous nuclear 
polarization, and $\Phi _{0}$ is the phase difference due to the nuclear 
polarization between the two paths at time $t = 0$~s. The lab-time 
dependence of the amplitude and phase of the combined spin packet are 
simulated from Eq. (3) and (4). The model gives good agreement to data 
with $P_u = 0.57$, $p_u = 0.45$, $p_l = 0.70$, $\Phi _0 = 4.5\pi $, and 
$\tau _n = 400$~s [Fig. 2 (c)]. For the analysis of the phase, we have 
subtracted an exponentially decaying component and an offset given by $\Phi 
_1 \exp \left( { - t / \tau _n } \right) + 0.75\pi $ with $\Phi _1 = 2.86\pi 
$ and $\tau _n = 400$~s, which is shown as a solid line in Fig. 2 (d), and 
the data after subtraction is shown in Fig. 2 (e). The simulation of the 
lab-time dependence of the phase is obtained from the same set of parameters 
used in Fig. 2 (c), and is shown in Fig. 2 (f). A good agreement is seen 
between the data and the model for the phase as well.

The parameters obtained in the analysis allow us to infer the spatial extent 
of the nuclear spin polarization. $\Phi _{1}$ gives a measure of the 
nuclear polarization in the upper arm, while $\Phi _{0}+\Phi _{1}$ 
gives a measure of that in the lower arm. The ratio of these two values is 
$\sim $2.5, indicative of significant nuclear polarization in the upper arm. 
This can be explained by electron spin diffusion, which has been seen to 
extend to $\sim $15 $\mu $m in a similar system. \cite{Stephens:2003} In 
addition, suppression of nuclear spin polarization at the center of a laser 
spot has also been observed previously, \cite{Stephens:2004} and this may 
be playing a role in keeping this ratio to a relatively small value. There 
may also be nuclear polarization arising from electron spins excited in the 
upper arm by tails of the DNP beam.

We note that the interference we observed is of net spin polarization of the 
electrons, and is not quantum interference 
\cite{Engel:2000,Yau:2002,Koga:2004} in the sense that 
the orbital part of the electron wave function has decohered. We also note 
that the Aharonov-Bohm effect 
\cite{Aharonov:1959,Webb:1985,Timp:1987} is not expected 
here as the magnetic field is applied in the plane of the sample.

In summary, we have demonstrated electron spin interferometry using a 
semiconductor ring structure. It takes advantage of the long lifetime of 
spin polarization relative to charge coherence, and may find applications in 
detecting magnetic field gradients. The measurements also provide insights 
into the effect of nuclear polarization on electron spin dephasing.

We thank G. Salis and acknowledge support from the DARPA and the DMEA.

\end{document}